\documentclass[
    ,final            
  ]
  {aipproc}

\usepackage{amssymb}

\layoutstyle{8x11double}

\renewcommand\XFMtitleblock{%
  \XFMtitle
  \let\XFMoldpar\par
  \def\par{\XFMoldpar\def\par{\space 
             (on behalf of the VERITAS Collaboration)\XFMoldpar}}%
   \XFMauthors
   \let\par\XFMoldpar
   \XFMaddresses
   \XFMabstract
   \vspace{5pt}%
   \XFMkeywords
   \XFMclassification
 }

\begin{document}

\title{VERITAS Studies of the Supernova Remnants \newline Cas A and IC 443}


\author{T. B. Humensky}{
  address={University of Chicago Enrico Fermi Institute, Chicago, IL, USA}
}

\begin{abstract}
VERITAS observed the supernova remnants Cassiopeia A (Cas A) and IC 443 during 2007, resulting in strong TeV detections of both sources. Cas A is a young remnant, and bright in both the radio and nonthermal X-rays, both tracers of cosmic-ray electrons. IC 443 is a middle-aged composite remnant interacting with a molecular cloud; the molecular cloud provides an enhanced density of target material for hadronic cosmic rays to produce TeV gamma rays via pion decay. The TeV morphology - point-like for Cas A and extended for IC 443 - will be discussed in the context of existing multiwavelength data on the remnants.
\end{abstract}

\maketitle


\section{Introduction}
Fermi acceleration in supernova remnants (SNRs) has long been considered as one of the likely explanations for the origin of cosmic rays (CRs) up to the knee at $\sim 3\cdot 10^{15}\ \textrm{eV}$. This scenario drives the observations of SNRs in the very-high-energy (VHE) gamma-ray band, where measurements of the morphology and energy spectrum, when combined with multiwavelength observations, hold promise to constrain models of diffusive shock acceleration and the potential to discriminate between gamma-ray production by leptonic and hadronic cosmic rays.  For example, the deep H.E.S.S. observations of RX J1713.7-3946 reveal an energy spectrum extending beyond $30\ \textrm{TeV}$, implying particle acceleration to well beyond $100\ \textrm{TeV}$\cite{2007A&A...464..235A}. Unfortunately, in many cases (including RX J1713.7-3946) uncertainty in the environment makes it difficult to discriminate conclusively between hadronic and leptonic scenarios for the gamma-ray emission.

Cassiopeia A (Cas A) is a young remnant, believed to be about 300 years old and at a distance of $3.4\ \textrm{kpc}$.  It has a diameter of 5 arcmin, making it a point source for TeV telescopes.  Cas A was first detected in TeV by HEGRA\cite{Aharonian2001a}, with a flux of $\sim 3.3\%$ Crab above $1\ \textrm{TeV}$.  The detection was confirmed by MAGIC\cite{Albert2007a} at $5.3\ \sigma$ in a 47-hour observation, yielding a flux above $1\ \textrm{TeV}$ of $(7.3 \pm 0.7_{stat} \pm 2.2_{sys}) \times 10^{-13}\ \textrm{cm}^{-2}\textrm{s}^{-1}$ and a spectral index of $2.3 \pm 0.2_{stat} \pm 0.2_{sys}$, consistent with the HEGRA results.  Cas A is an interesting TeV SNR because of its low age, detailed observations across all wavelengths, and relatively simple local environment, making it attractive for constraining models of particle acceleration in shocks.

IC 443, by contrast, is one of the classic examples of a SNR interacting with a molecular cloud - placing it in a complex environment and greatly affecting its evolution.  While the environment makes modeling difficult, TeV emission coincident with the site of the shock/cloud interaction arguably provides a smoking gun for the acceleration of hadronic cosmic rays.  IC 443 is about $1.5\ \textrm{kpc}$ away and has a diameter of $\sim 0.75\textrm{\textdegree}$, with an appearance of two half-shells.  One is expanding to the northeast and interacting with an H \i\ region, and the other is expanding to the southwest, where it is interacting with a giant molecular cloud at several points.  The cloud has a mass of $\sim 10^4\ \textrm{M}_{\odot}$\cite{2003mora.meet..323B}, mostly along the line of site to IC 443; maser emission has been observed in this region\cite{Claussen1997a}.  The age of IC 443 is unclear; Troja et al. place it at $\lesssim 10\ \textrm{kyr}$, possibly around $4\ \textrm{kyr}$\cite{Troja2008a}.  The PWN CXOU J061705.3+222127 is located at the southern edge of the remnant\cite{Olbert2001a,Bocchino2001a,Gaensler2006a}.  While pulsed emission has not yet been detected, the inferred $\dot{\textrm{E}}$ has been estimated indirectly at $\sim 10^{36}\ \textrm{erg s}^{-1}$\cite{Olbert2001a,Bocchino2001a} and $\sim 5 \times 10^{37}\ \textrm{erg s}^{-1}$\cite{Gaensler2006a} - a large range, but suggesting a fairly energetic pulsar.  Both MAGIC and VERITAS have reported TeV emission from the central region of the remnant, coincident with the maser emission\cite{Albert2007a,2007arXiv0709.4298V}.  An EGRET GeV source also overlaps the remnant\cite{1995A&A...293L..17S,Esposito1996a}.

VERITAS observed both Cas A and IC 443 during 2007, and the results of those observations are discussed below.

\section{Observations and Analysis}

VERITAS is an array of four 12-m-diameter Cherenkov telescopes located at the base camp of the Whipple Observatory (111.0\textdegree\ W, 31.7\textdegree\ N, elevation $1268\ \textrm{m}$).  Each telescope is equipped with 499-pixel camera providing a 3.5\textdegree\ field of view.  Signals are buffered for $32\ \mu\textrm{s}$ in 500-Msps FADCs.  A camera is triggered when three adjacent pixels each record a signal above $50\ \textrm{mV}$ ($\sim 4.5\ \textrm{p.e.}$) within a $13\textrm{-ns}$ window.  The array is triggered and the DAQ read out when at least two cameras trigger within $100\ \textrm{ns}$.   The typical array trigger rate is $\sim 270\ \textrm{Hz}$ near zenith.

Both Cas A and IC 443 were observed in wobble mode, with the telescope pointing offset by 0.5\textdegree\ from the target direction for roughly equal amounts of time in each of four equally spaced directions.  Cas A was observed for 25 hours in October/November, 2007 with four telescopes. After data quality selection (for rate stability and good weather) and correction for deadtime, an exposure of 20.3 hours remains with an average zenith angle of 30\textdegree.  

IC 443 was observed during two periods: with three telescopes in February/March, 2007 (during the commissioning phase of the array) for 20 hours; the PWN was the target for these observations.  An additional 25 hours of observations were taken in October/November, 2007 using the full four-telescope array and targeting the location of the emission observed previously, 06 16.9 +22 33.  After quality selection and deadtime correction, the data set has a livetime of 37.1 hours and an average zenith angle of 18\textdegree.

The data analysis is described more fully in~\cite{Daniel2007a}.  In brief, pixel gains are flatfielded user laser flashes recorded during nightly calibration runs. Charges are integrated over a seven-sample (14 ns) window.  A two-pass cleaning algorithm with thresholds of 5 and 2.5 times the pedestal RMS is used; isolated pixels which survive the image cleaning are then eliminated.  The images are parameterized according to the Hillas prescription~\cite{1985ICRC....3..445H}, and then the shower direction and core location is determined.  Lookup tables provide expected values for the image width, length, and energy, and are used to construct the gamma-hadron separation parameters Mean Scaled Width (MSW) and Length (MSL), and an estimate of the shower energy.  Images are included in the reconstruction if they have at least five pixels, if the centroid is less than 1.43\textdegree\ from the center of the camera, and if the image size (sum of the pixel charges) is at least 400 d.c. ($\sim 75\ \textrm{p.e.}$).  Gamma-hadron separation is applied by requiring at least two good images per event, cutting events in which only T1 and T4 have good images (due to their short baseline, which leads to relatively poor reconstruction), and requiring 0.05 < MSW < 1.16 and 0.05 < MSL < 1.36.  For generating maps, an integration (smoothing) radius of 0.115\textdegree\ is used.  These cuts have been optimized for weak sources with a Crab-like spectrum.  The background is estimated using the ``ring background model'' (see, e.g., \cite{Berge2007a}).

The IC 443 field contains two bright stars, including one that is $\sim 0.5\textrm{\textdegree}$ from the center of the remnant.  The stars require that several pixels be turned off during observations to prevent damage to the PMTs, suppressing the system acceptance for showers that originate in their direction.  This is dealt with in the analysis by excluding regions around the stars from the background estimation for other parts of the map (though this leaves a deficit at the location of the star in the map).  To confirm that the stars are not unduly affecting the analysis, the results are compared with those derived from an independent analysis chain using a template background model.

\section{Cassiopeia A}

\begin{figure}
  \includegraphics[height=.3\textheight]{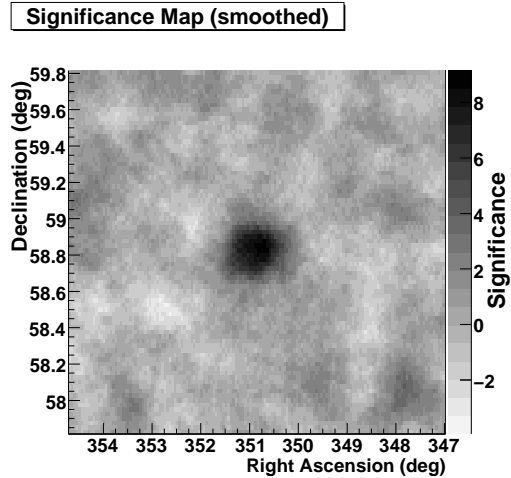}
  \caption{Cas A significance map.}\label{CasASigMap}
\end{figure}

\begin{figure}
  \includegraphics[width=.45\textwidth]{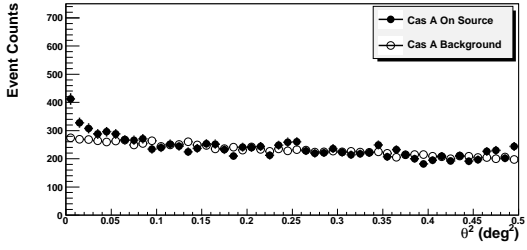}
  \caption{Cas A theta-squared distribution.}\label{CasAThetasq}
\end{figure}

Figure~\ref{CasASigMap} shows the significance map for the Cas A field and Figure~\ref{CasAThetasq} shows the distribution of events as a function of the square of the angular distance from the source location.  Clear peaks are seen at the location of Cas A, with a significance of $9.8\ \sigma$ and a flux (determined by comparing to the rate of gamma rays from the Crab Nebula) of $\sim 3\ \%$ Crab, consistent with the HEGRA and MAGIC results.  The emission is consistent with a point source within the instrument's angular resolution, and the spectral analysis is under way.  

\section{IC 443}

\begin{figure}
  \includegraphics[height=.3\textheight]{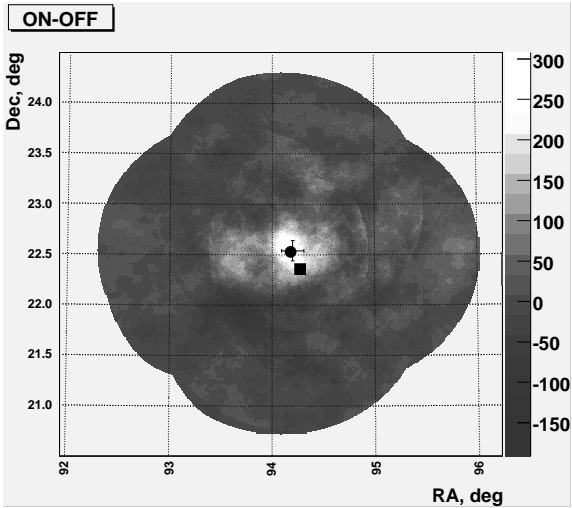}
  \caption{Excess map for IC 443 using a template background model from an independent analysis chain. Note that the RA axis is reversed with respect to Figure~\ref{IC443MWL}.  The cross indicates the VERITAS centroid and uncertainty, the black circle indicates the MAGIC centroid, and the black square indicates the location of the PWN.}\label{IC443Template}
\end{figure}

\begin{figure}
  \includegraphics[bb=0 50 567 490,clip,width=.58\textwidth]{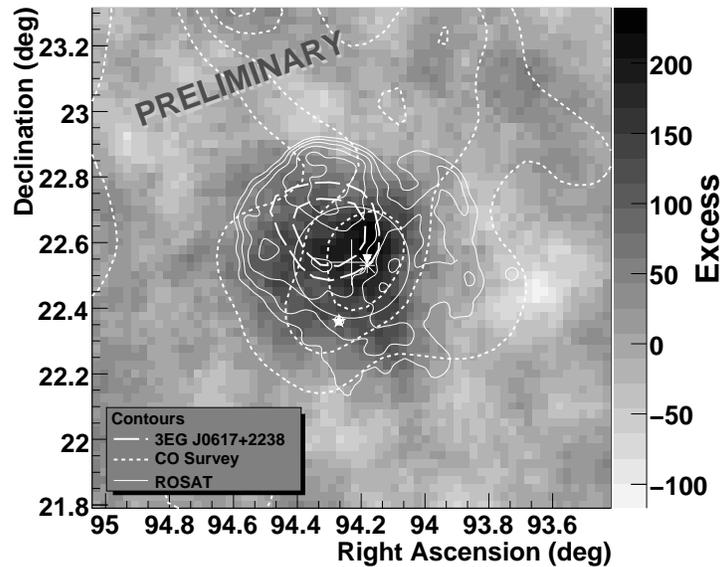}
  \caption{IC 443 excess map in a multiwavelength context.  The color scale indicates the TeV excess rate (arbitrary units).  The VERITAS centroid is indicated by the cross, with the uncertainty in position indicated by the lengths of the arms.  The one-sigma extension is indicated by the white circle.  The MAGIC centroid is indicated by a white star, immediately below the triangle that marks the location of maser emission.  The PWN is indicated by a filled white star.  Contours are overlayed indicating CO emission (short dashed lines), X-rays (solid lines), and GeV (long diashed lines).}\label{IC443MWL}
\end{figure}

Figures~\ref{IC443MWL} and~\ref{IC443Template} show the excess maps from the ring background model and template analyses, respectively.  They agree well on the centroid and extension of the excess.

\begin{figure}
  \includegraphics[width=.45\textwidth]{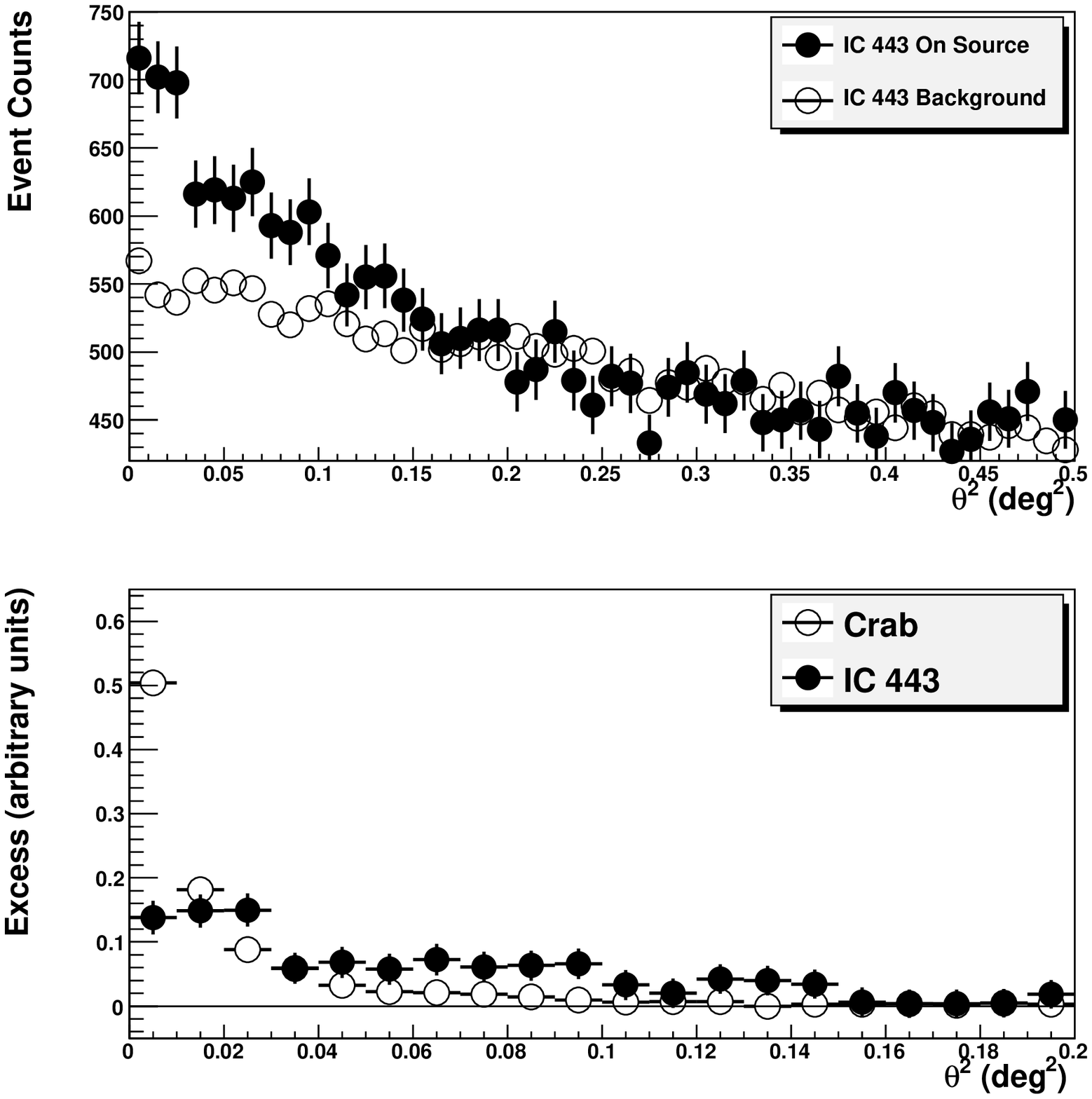}
  \caption{(top) IC 443 theta-squared distribution for on-source (filled circles) and background (open circles) data. (bottom) IC 443 theta-squared excess compared to the Crab Nebula, zoomed in to the region of excess.}\label{thetasqIC443}
\end{figure}

Figure~\ref{thetasqIC443} shows the distribution of theta-squared for IC 443, compared with the same distribution for observations of the Crab Nebula.  The extended nature of the IC 443 emission is clear.  Fitting a 2-d gaussian to the uncorrelated excess map yields a centroid location of 06 16.9 +22 32.4 $\pm 0.03\textrm{\textdegree}_{stat} \pm 0.07\textrm{\textdegree}_{sys}$ and an extension of $0.17\textrm{\textdegree} \pm 0.02\textrm{\textdegree}_{stat} \pm 0.04\textrm{\textdegree}_{sys}$, after accounting for the intrinsic TeV PSF.  This extension is not inconsistent with MAGIC's report of a point-like source, given the difference in angular resolution between the two instruments and the much deeper VERITAS observation.

Figure~\ref{IC443MWL} shows the TeV emission from IC 443 in a multiwavelength context.  It is striking that the TeV emission correlates well in position with the densest concentration of matter and with the maser emission.  This spatial coincidence suggests that the observed gamma radiation is produced by hadronic cosmic rays accelerated in the remnant's shock, either in a scenario where a portion of the SNR evolves in the molecular cloud\cite{2008ApJ...675L..21Z} or in a scenario where cosmic rays leaking out of the SNR have propagated ahead of the shock and into the cloud\cite{2003mora.meet..323B,2008MNRAS.387L..59T}.  The recent AGILE detection of GeV emission~\cite{Tavani2008a} shifts the GeV centroid away from the TeV emission, and suggests that they are two independent sources.

The PWN CXOU J061705.3+222127, with an offset of $\sim 0.2\textrm{\textdegree}$ from the TeV centroid, is an alternate candidate as the source of TeV emission\cite{2008MNRAS.385.1105B}.  However, as \cite{2008MNRAS.387L..59T} argues, it is difficult in this scenario to reconcile either the displacement of the EGRET source with respect to the PWN or the TeV morphology with the distribution of target photons available to drive inverse Compton scattering.

\section{Conclusions}
VERITAS has made strong detections of both the Cassiopeia A and IC 443 supernova remnants.  Cas A is observed to have a flux of $\sim 3\ \%$ Crab and appears pointlike within the instrument's resolution.  These results are consistent with previous observations.  

The emission from IC 443 is extended, yielding a sigma of $\sim 0.17\textrm{\textdegree}$ when fit with a two-dimensional gaussian.  It has an integral flux of $\sim 3\ \%$ Crab above $200\ \textrm{GeV}$.  The TeV emission is coincident with both the densest part of the molecular cloud that IC 443 is interacting with and the maser emission observed within that cloud, strongly suggesting that the gamma rays are produced by the interaction of hadronic cosmic rays with cloud material.  An alternative explanation in terms of an association with the PWN CXOU J061705.3+222127 is also possible but seems less likely.



\begin{theacknowledgments}
This research is supported by grants from the U.S. Department
of Energy, the U.S. National Science Foundation,
and the Smithsonian Institution, by NSERC in
Canada, by PPARC in the UK and by Science Foundation
Ireland.
\end{theacknowledgments}


\bibliographystyle{aipproc}   

\bibliography{HumenskySNRsVERITAS_v1}

\IfFileExists{\jobname.bbl}{}
 {\typeout{}
  \typeout{******************************************}
  \typeout{** Please run "bibtex \jobname" to optain}
  \typeout{** the bibliography and then re-run LaTeX}
  \typeout{** twice to fix the references!}
  \typeout{******************************************}
  \typeout{}
 }

\end{document}